# Terahertz-Field-Induced Ferroelectricity in Quantum Paraelectric SrTiO$_3$

Xian Li[1], Tian Qiu[2], Jiahao Zhang[2], Edoardo Baldini[3], Jian Lu[1], Andrew M. Rappe[2], Keith A. Nelson[1*]

[1]Department of Chemistry, Massachusetts Institute of Technology, Cambridge, MA 02139, USA

[2]Department of Chemistry, University of Pennsylvania, Philadelphia, PA 19104-6323, USA

[3]Department of Physics, Massachusetts Institute of Technology, Cambridge, MA 02139, USA

*Correspondence to: kanelson@mit.edu

"Hidden phases" are metastable collective states of matter that are typically not accessible on equilibrium phase diagrams. These phases can host exotic properties in otherwise conventional materials and hence may enable novel functionality and applications, but their discovery and access are still in early stages. Using intense terahertz electric field excitation, we show that an ultrafast phase transition into a hidden ferroelectric phase can be dynamically induced in the quantum paraelectric SrTiO$_3$. The induced lowering in crystal symmetry yields dramatic changes in the phonon excitation spectra. Our results demonstrate *collective coherent control* over material structure, in which a single-cycle field drives ions along the microscopic pathway leading directly to their locations in a new crystalline phase on an ultrafast timescale.

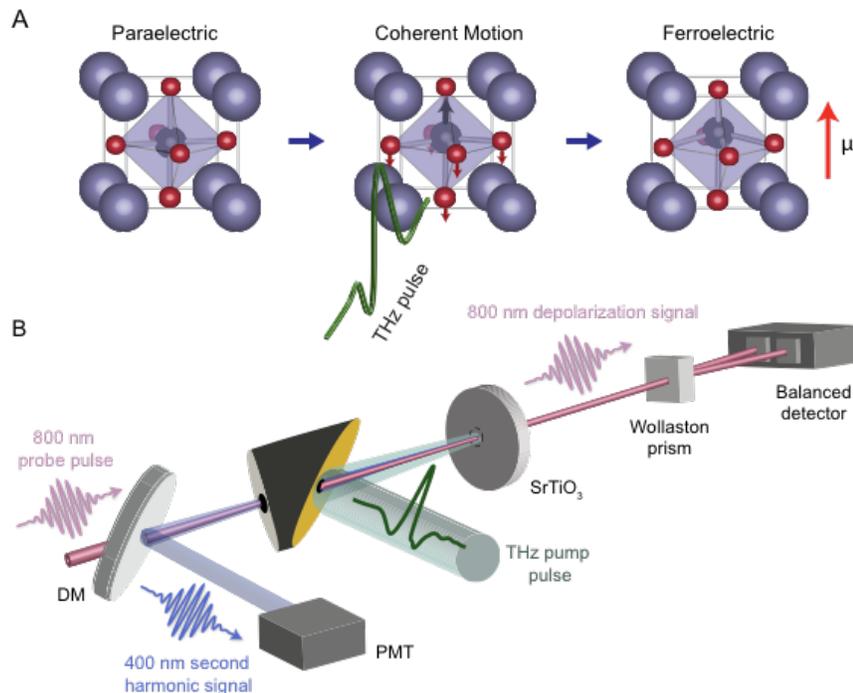

**Fig. 1. Hidden ferroelectric phase accessed through THz field excitation.** (A) Collective coherent control over material structure. A single-cycle terahertz-frequency electric field moves all the ions it encounters toward their positions in a new crystalline phase. In strontium titanate, the initial high-symmetry configuration around each Ti$^{4+}$ ion has no dipole moment and the crystal is paraelectric. The incident field drives the "soft" lattice vibrational mode, moving the ions along the directions indicated into a lower-

symmetry geometry with a dipole moment. Long-range ordering of dipole moments in the same direction yields a ferroelectric crystalline phase. (B) Experimental setup. THz field-induced lowering of the SrTiO$_3$ crystal symmetry is observed using 800-nm probe pulses that are partially depolarized (terahertz Kerr effect, or TKE) and which are partially converted to the second harmonic frequency (THz field-induced second harmonic, or TFISH). DM: dichroic mirror; PMT: photomultiplier tube.

In recent years, significant advances have been made in the search for materials with complex multi-phase landscapes that host photoinduced metastable collective states, or "hidden" phases. These phases are rarely accessible on equilibrium phase diagrams and may persist long after the external stimuli that induced them are over. Recent experiments (*1-6*) have illustrated some of the possibilities and expanded our understanding of nonequilibrium material properties and dynamics. In some cases, ultrafast resonant excitation of crystal lattice vibrations (phonons) has played the key role in reaching hidden metallic and superconducting phases (*1, 2*). Here we extend this capability in the discovery of a hidden ferroelectric (FE) phase in the paradigmatic material SrTiO$_3$ (STO). We accessed the hidden phase by selectively exciting the soft phonon mode that serves as a collective reaction coordinate along which ions move from their initial positions toward their positions in the new phase (see Figure 1). The resulting ultrafast control over ferroelectricity may find rich applications in memory devices (*7*), STO-based heterostructures (*8*) and high-$T_c$ superconductivity (*9, 10*). In other recent experiments (*11*), ionic displacements along soft-mode coordinates have been driven through nonlinear coupling between the soft modes and other phonon modes that were excited by long-wavelength infrared pulses. In the present case, we excite the soft mode directly, using a terahertz (THz) light field to move the ions into their positions in the incipient crystalline phase. This case was foreshadowed by molecular dynamics (MD) simulations of THz field-induced switching between different FE domain orientations, a closely related type of "collective coherent control" (*12*).

STO is a widely-used dielectric material that has a cubic perovskite structure at room temperature. Many members of this crystal family, e.g. PbTiO$_3$, undergo transitions into FE phases in which the transition-metal ions occupy positions that are displaced from the unit cell center so that the material has a macroscopic electric polarization. The collective pathway between the cubic, paraelectric phase and the FE phase involves motions of the ions along the soft phonon coordinate illustrated in Fig. 1A. In contrast, upon reduction of the temperature to 105 K, STO undergoes an antiferrodistortive structural phase transition into a second paraelectric phase of tetragonal symmetry (*13, 14*). Further cooling reveals mode softening (decrease in frequency $\omega$) in the usual Curie-Weiss form $\omega \propto (T - T_c)^{1/2}$ with critical temperature $T_c$ = 36 K (*15*), but at such a low temperature the zero-point quantum uncertainties in ion positions prevent long-range FE ordering of their locations. Thus STO is a textbook example of a so-called quantum paraelectric (QPE) phase (*15*), in which dipole correlation lengths do not extend beyond nanometer length scales ("polar nanoregions", or PNRs) (*16*). Recently, studies have shown that the QPE state in STO is a result of a more complex competition among three driving forces (*17, 18*), *i.e.* quantum fluctuations, antiphased structural distortions (rotations of neighboring oxygen octahedra in opposite directions), and ferroelectric ordering. As a result, even subtle perturbations such as $^{18}$O isotope substitution (*19*) are able to turn STO ferroelectric.

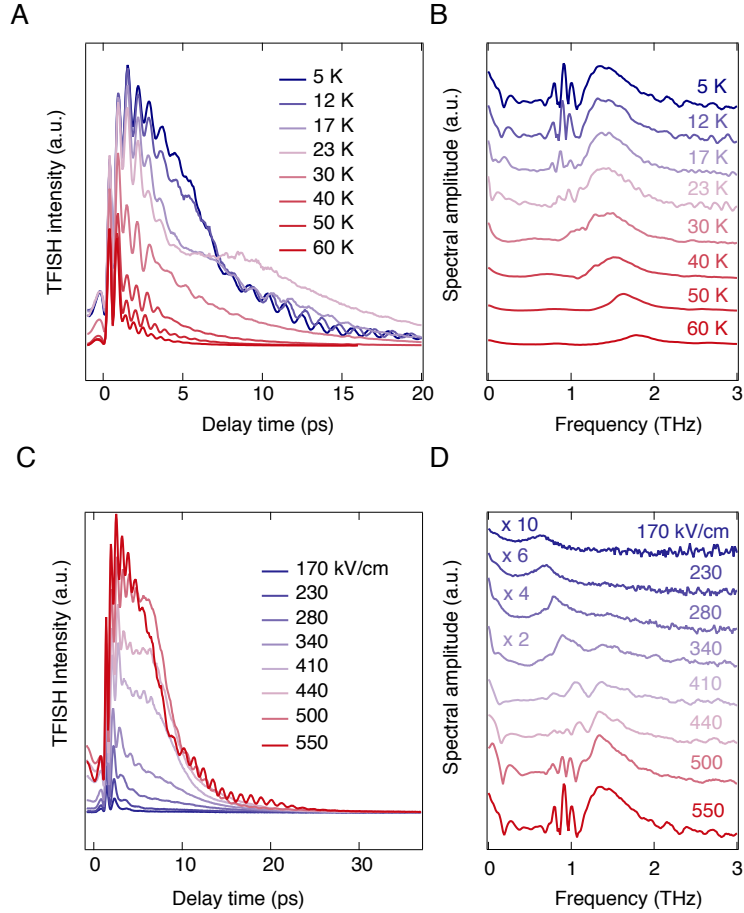

**Fig. 2. STO symmetry breaking measured by optical second harmonic generation (TFISH).** (A) Temperature-dependent TFISH signals recorded at 550 kV/cm field amplitude from STO and (B) their Fourier transforms. The ferroelectric soft mode is observed above 30 K, and new phonon peaks as well as non-oscillatory signals appear at lower temperatures. (C) THz field-strength-dependent TFISH signals measured at 5 K and (D) their Fourier transforms. Signals at low field strengths are magnified by the amounts indicated in the figure for better visibility. Dramatic changes in the non-oscillatory signal components and the phonon spectra occur when the THz field level is increased above 340 kV/cm. The numerical first derivatives of the time-domain signals were calculated before Fourier transformation to reduce the relative amplitude of the non-oscillatory components.

In this Report, we show that intense coherent THz excitation of the FE soft modes in STO can lead to highly nonlinear phonon responses that overcome the quantum fluctuations and yield clear signatures of an ultrafast QPE-to-FE phase transition. The observed signals reveal a dramatic rise in ferroelectric ordering and restructuring of phonon spectra beyond a threshold THz field strength, indicating the emergence of the collective FE phase.

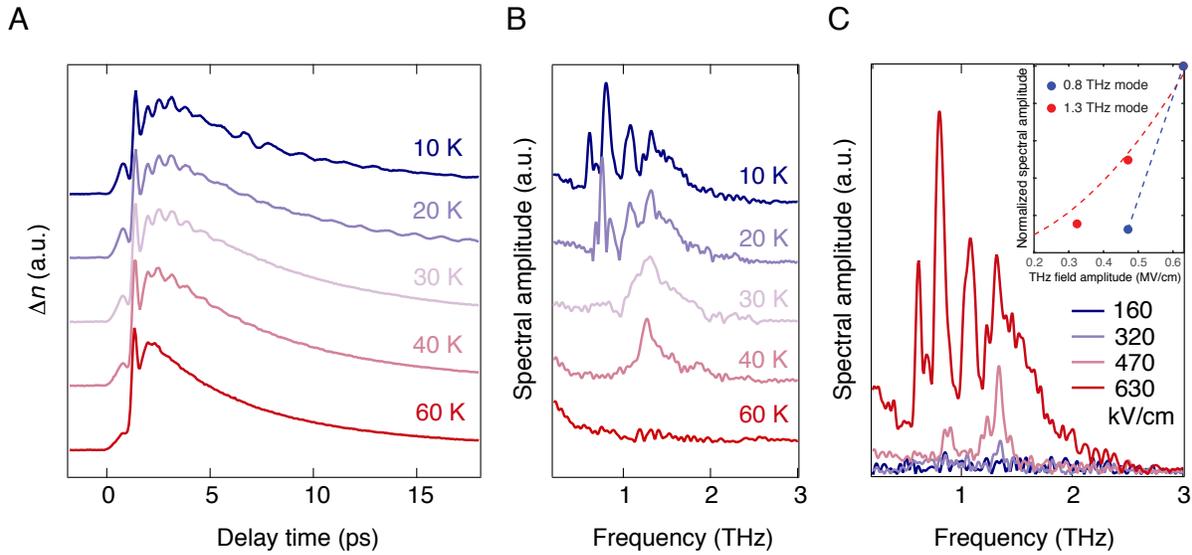

**Fig. 3. Strongly nonlinear phonon responses appear in the low-symmetry STO phase.** (A) Temperature dependence of THz-induced optical depolarization (TKE) signals recorded with 630 kV/cm THz pump field amplitude and (B) their Fourier transforms. The numerical first derivatives of the time-domain signals were calculated before Fourier transformation to reduce the relative amplitude of the non-oscillatory components. At temperatures 60 K and above (see Supplement), no oscillatory signal is observed after THz excitation. The 1.3 THz peak and additional low-frequency modes appear at low temperatures. (C) THz field strength dependence of the TKE spectra at 10 K. New peaks grow in sharply as the THz field level is increased from 470 to 630 kV/cm. Inset: quadratic fit to the 1.3 THz $A_{1g}$ mode. The 0.8 THz mode shows faster than quadratic scaling in the THz field strength.

Two complementary experiments were carried out with single-cycle THz pump pulses and time-delayed optical probe pulses, as illustrated in Fig. 1B. THz-field-induced second-harmonic (TFISH) generation spectroscopy (*20*) was conducted to observe signals that arise from symmetry breaking due to coherent soft mode lattice vibrational motion away from the initially centrosymmetric structure of the QPE phase. THz field-induced optical birefringence (THz Kerr effect, or TKE) spectroscopy (*21*) was performed to characterize Raman-active phonon responses that were driven nonlinearly by the THz-initiated soft mode lattice vibrations. Figure 2 shows TFISH measurement results from STO and their Fourier transforms at several temperatures and THz field amplitudes. At temperatures above 30 K, a single mode that softens with decreasing temperature consistent with the ferroelectric soft mode (*14*) is observed. The coherent vibrational displacements in either direction break the symmetry, resulting in optical second harmonic signals that oscillate at twice the soft mode frequency. There is also a non-oscillatory signal component due to THz-induced orientation of dipoles, whose decay becomes slower as *T* is reduced due to the increasing correlation lengths of the PNRs (*22*). In the QPE phase at *T* < 36 K, two features in the signals change dramatically at high THz field amplitudes. First, the non-oscillatory signal component grows in a highly nonlinear fashion as a function of the field strength. This indicates a dramatic growth in the extent of steady-state (non-oscillatory) dipole ordering (*23*). Second, additional phonon signatures appear with amplitudes that also increase in a highly nonlinear fashion with THz field strength. These features reveal additional ionic displacements that take place as the FE crystal structure is formed. At soft mode amplitudes sufficient to reach the new phase, collective displacements of other phonon modes (coupled nonlinearly to the soft mode) are induced. The THz-induced ordered structure is noncentrosymmetric, so oscillations about this structure produce changes in the second harmonic signal level that oscillate at the phonon frequencies, not twice the frequencies. It is noteworthy that the three distinct low-frequency peaks in the TFISH response at high field strength harden gradually as *T* is reduced (shown in Fig. S6 of the Supplement), as is known to occur for the Raman-active phonon modes associated with the STO phase transition at 105 K. It is likely that we are observing these modes, with frequencies altered slightly due to the nonequilibrium transient crystal structure in which we are observing them, and that their sharp onset at high fields indicates their displacements associated with the FE crystal structure. We also observe a broad phonon feature at 1.3 THz whose frequency does not appear to change with temperature and whose signal strength does not increase as sharply as the lower-frequency peaks. We believe this behavior is consistent with a Raman-active $A_{1g}$ mode (*24*) that is coupled anharmonically to the soft mode, not associated specifically with the FE phase transition. Similar nonlinear coupling has been observed in room-temperature STO using femtosecond x-ray diffraction (*25*).

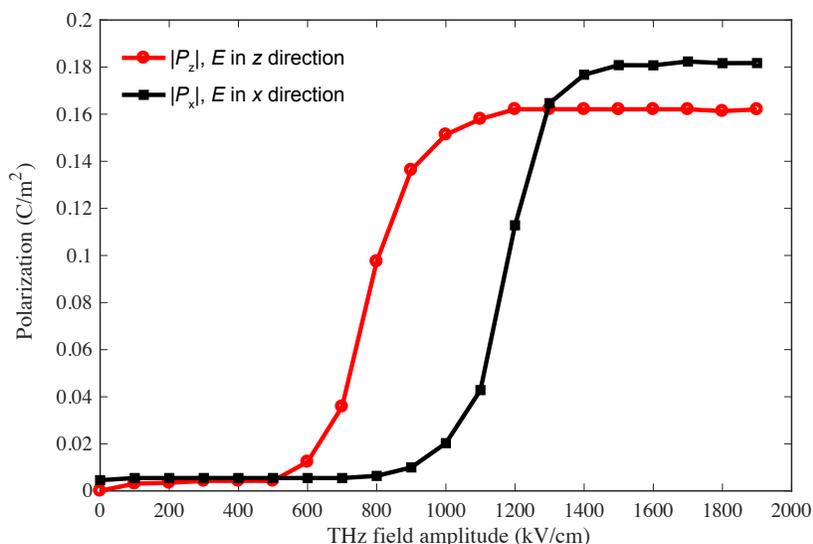

**Fig. 4. MD simulation of the global polarization in STO after THz field excitation**. Global polarization versus amplitude of the THz electric field along different crystallographic axes.

Figure 3 shows TKE data recorded at several sample temperatures and THz field strengths. At high temperatures, as shown in Fig. 3A, only non-oscillatory signals are observed. Unlike such signals in the TFISH data, these signals show only weakly *T*-dependent decay kinetics and they do not increase dramatically as functions of either temperature or THz field amplitude (see Fig. S4 in Supplement.) We believe they are associated with dipole alignment rather than ferroelectric orientation. We also observe the $A_{1g}$ mode at 1.3 THz which increases quadratically with THz field strength, indicating ordinary anharmonic coupling to the FE soft mode as suggested above. By far most striking is the emergence of several low-frequency phonon features whose strengths depend in a highly nonlinear fashion on the THz field strength, clearly similar to what we observed in TFISH measurements. We conclude from all the experimental evidence that at sufficiently large soft mode amplitudes, an ultrafast ferroelectric phase transition is triggered. The strong non-oscillatory TFISH signals reveal the associated increase in FE ordering. The modes that grow in sharply as the THz field amplitude is increased reveal significant collective displacements of ions along multiple vibrational modes that are coupled nonlinearly to the soft mode and also reveal the change in lattice symmetry. It has been suggested that excitation of the Raman modes may provide constructive feedback to the FE soft mode that drives them by disrupting the balance between antiferrodistortive and ferroelectric structural distortions (*17, 18, 26*), thereby dynamically destabilizing the paraelectric ground state on a multidimensional energy landscape.

In order to reach a clearer understanding of THz-induced effects, we conducted classical MD simulations for a supercell of 20×20×20 unit cells in an isothermal-isobaric ensemble, with the interatomic interaction described by the bond-valence model (see Supplement for the simulation details). For each MD simulation, the system was first relaxed for 100 ps to reach equilibrium at 5 K, and then a Gaussian-profile electric field pulse with full-width at half-maximum duration of 0.66 ps was applied in either the *z* or *x* crystallographic direction. Trajectories of the system were collected for 20 ps after the electric field returned to zero. To evaluate whether the electric field could induce ferroelectricity in STO, simulations were performed with different field amplitudes, and in each case the global polarization of the system was calculated from the collected trajectories. As shown in Fig. 4, the applied electric field has little influence on the global polarization until the maximum field reaches a certain threshold, which is around 800 kV/cm in the *z* direction and 1100 kV/cm in the *x* direction. Note that the experimental crystal in the tetragonal phase below 105 K is likely not single-domain, so the THz field polarization lies along the unique *z* direction in some domains and the *x* direction in others. The comparison between simulated and experimental field amplitudes is therefore semi-qualitative. The important result of the simulations is the confirmation that a single-cycle THz field can induce a long-lived FE polarization when the field is above a threshold level on the order of 1 MV/cm.

The experimental data and molecular dynamics simulations together demonstrate a THz-induced ultrafast quantum-paraelectric-to-ferroelectric phase transition in STO. The THz field drives the soft mode, and additional coupled-mode displacements occur to reach the FE structure. Our results demonstrate collective coherent control of material structure that may be applicable to a wide range of classical and quantum phase transitions in which soft phonon modes play key roles in the collective structural transformations.

**Acknowledgments:** We acknowledge A. Steinbacher, Y. Wang and E. Demler for stimulating discussions. **Funding:** The work at MIT was supported in part by U.S. Department of Energy, Office of Basic Energy Sciences, under Award No. DE-SC0019126. T. Q. was supported by the National Science Foundation, under grant number CHE-1808202. J. Z. was supported by the National Science foundation, under grant DMR-1719353. A. M. R. was supported by the Office of Naval Research, under grant N00014-17-1-2574. The authors also acknowledge computational support from the HPCMO of the DoD. E. B. acknowledges support by the Swiss National Science Foundation under fellowship P2ELP2-172290; **Author contributions:** K.A.N. conceived the project and the experiments together with X.L. and J.L. The time-resolved THz set-up was built by X.L., who performed the TFISH and TKE measurements and analyzed the data with support from E.B. T.Q., J.Z. and A.M.R. provided MD calculations of the pump thresholds for the THz-driven ferroelectricity. The manuscript was written by K.A.N., X.L. and E.B., with input from all authors; **Competing interests:** Authors declare no competing interests; and **Data and materials availability:** All data is available in the main text or the supplementary materials.

**Supplementary Materials:**

Materials and Methods

Supplementary Text

Figs. S1 to S7

Table S1

References (27-43)

# Supplementary Materials for

**Terahertz-Field-Induced Ferroelectricity in Quantum Paraelectric SrTiO$_3$**

Xian Li, Tian Qiu, Jiahao Zhang, Edoardo Baldini, Jian Lu, Andrew M. Rappe, Keith A. Nelson*

Correspondence to: kanelson@mit.edu

**Materials and Methods**

The samples were Verneuil-grown, 10×10×0.5 mm³ SrTiO₃ single crystals with epi-polished (100) surfaces obtained from MTI Corp. STO has a cubic perovskite lattice structure (space group $Pm\bar{3}m$) at room temperature and undergoes an antiferrodistortive structural transition to a tetragonal structure (space group *I4/mcm*) at 105 K (*13*), which leads to a unit cell doubling in the material. In the paraelectric phase, four modes exist for STO in the relevant frequency range: the IR active $A_{2u}$, $E_u$ modes and the Raman-active $A_{1g}$ and $E_g$ modes (*27*).

Our pulsed laser is a 1 kHz repetition rate Ti:Sapphire amplifier (Coherent Legend Elite Duo) seeded by an 80 MHz Ti:Sapphire oscillator (Coherent Vitara T) with an output power of 13 W at 800 nm and 35 fs pulse duration. 7 W of the power is split as the input to our setup and 95% of that input is used to generate the THz field. Single-cycle THz pulses are generated from a LiNbO₃ crystal by optical rectification using the tilted-pulse-front technique (*28, 29*). The 800-nm pump pulse is modulated at 500 Hz by a mechanical chopper to isolate the signals due to THz excitation. A 4-*f* imaging system containing two 90° off-axis parabolic mirrors is used to collimate and focus the THz field onto the sample in the cryostat. The remaining 5% laser input is attenuated and used as the optical probe in our experiments. The THz field strength is measured through field-induced depolarization of 800-nm light in a 100 μm GaP electro-optic crystal (i.e. using electro-optic sampling (EOS)) and can be varied by a pair of wire-grid polarizers (WGPs). The THz peak field strength in free space was measured to be 550 kV/cm and 630 kV/cm in the TFISH and TKE experiments, respectively. The THz field temporal profile is measured using EOS in a 300 μm GaP crystal after attenuation using the WGPs to avoid THz saturation of the EOS crystal. The setup is continuously purged with dry air to eliminate THz absorption due to water vapor in the air. The measured time-domain waveform of the THz electric field and its Fourier transform are shown in Fig. S1.

**Supplementary Text**

<u>TFISH Experiment Details</u>

In TFISH experiments, the THz-induced second harmonics (SH) of a normal-incidence 800 nm probe pulse was measured in reflection geometry using a photomultiplier tube. The second harmonic of the reflected probe was isolated through high-pass filters. Second harmonic generation is normally forbidden in centrosymmetric materials. In TFISH, the THz-induced responses (in STO, the soft mode displacements and associated dipole moments) break the symmetry and allow optical second harmonic generation to occur. Since SH generation is second-order in the incident optical field, and the soft mode displacements (at moderate field strengths) are linear in the THz field amplitude, the TFISH signal is third-order altogether in incident field amplitudes (The FE phase transition that occurs at strong THz fields is a highly nonlinear response, so the associated non-oscillatory signal and low-frequency phonon signals are higher than third-order.) Note that due to the use of pump chopping, any static SH signals that are not due to THz excitation are excluded. As a result, TFISH enables background-free detection that is highly sensitive to polar mode excitation (*30*). THz-induced phonon oscillations about the initial centrosymmetric crystalline structure break the symmetry during displacements in either direction, so time-dependent TFISH signal shows oscillations at twice the phonon frequency. However, when there is a steady-state, non-oscillatory breaking of inversion symmetry due to the THz-induced FE phase transition, phonon oscillations about the noncentrosymmetric structure modulate the TFISH signal at the fundamental frequency, and so the modes appear at their fundamental frequencies in the

TFISH spectra. The full time-domain TFISH temperature and field dependence data and their Fourier transforms are shown in Figs. S2 and S3. Note that weak TFISH signals at negative time delays, i.e. when the optical probe pulse precedes the THz pump pulse, arise because of a weak double internal reflection of the probe pulse inside the STO crystal. At low temperatures, the TFISH signals persist longer than the ~7.5-ps round-trip time of the probe pulse in STO, leading to the artifact near time zero.

TKE Experiment Details

The heterodyne TKE experiment setup is similar to that described earlier (*21*). The optical probe pulses traveled collinearly with the THz pulses in a transmission geometry and were polarized at 45º relative to the THz field polarization. The THz-induced optical birefringence $\Delta n$ was measured through the depolarization of the probe field using a balanced detection scheme. In the presence of inversion symmetry, system responses that are linear in the THz field are excluded by symmetry because lattice displacements that are IR (including THz) active are Raman inactive, which means that the refractive index does not change (to first order) with displacement. Thus soft mode displacements that are linearly proportional to the THz field amplitude do not change the refractive index (to first order) and do not give rise to TKE signals. Responses that are second-order in the THz field do appear. The full TKE time-domain temperature and field dependence data are shown in Figs. S4 and S5. The TKE signal can be decomposed into three components: an instantaneous electronic response that accounts for the initial spike in the TKE signal when the THz field arrives and is proportional to the square of the THz field profile; an orientational response due to THz-induced dipole alignment and its subsequent decay; and vibrational responses that are due to THz excitation of phonons as described in the main text. The penetration depth of the THz field in STO is on the order of microns at our THz frequencies due to the strong THz absorption (*14*). The large THz refractive index results in a large velocity mismatch (*21*) between the THz pulse and the optical probe pulse, but this does not lead to substantial timing variations over the short length in which the THz field strength is substantial. In both the TFISH and TKE experiments, only the signals after the main THz electronic response are included in the Fourier transforms, and numerical time-derivatives of the signals are calculated (i.e. signals from successive delay times are subtracted and the differences are divided by the time difference of 0.067 ps) before Fourier transformation to reduce the low-frequency backgrounds due to the non-oscillatory components of the signals.

Consideration of the possible origins of the 1.3 THz mode

The observed 1.3 THz peak agrees well with the previous measurements for the Raman $A_{1g}$ mode (*24, 31*). An assignment to two-quantum signals from the IR-active soft mode is ruled out because the observed 1.3 THz mode doesn't exhibit a clear frequency shift with temperature. Note that the 1.3 THz peak can be observed at a relatively high temperature of 40 K when STO is still in the conventional paraelectric phase. This implies that the Raman mode can be excited by THz fields even when STO is centrosymmetric, and this agrees with the nonlinear phononics interpretation of the excitation mechanism (*26*), i.e. the 1.3 THz mode is driven through anharmonic coupling to the soft mode.

Bond-valence model

The bond-valence model is an empirical force field for molecular dynamics (MD) simulation based on bonds between neighboring ions. The model contains five energy terms: Coulomb interaction, the short-range repulsive energy, the bond-valence energy, the bond-valence vector energy, and the angle potential. The total energy of the model is written as (*32*):

$$\begin{aligned} E_{\text{total}} &= E_c + E_{ij} + E_{bv} + E_{bvv} + E_{\text{angle}} \\ &= \sum_{i<j} \frac{q_i q_j}{r_{ij}} + \sum_{i<j} (\frac{B_{ij}}{r_{ij}})^{12} + \sum_i S_i (V_i - V_i^0)^2 \\ &+ \sum_i D_i ((\widehat{W}_i)^2 - (\widehat{W}_i^0)^2)^2 + \sum_{\text{oxygen}} k_i (\theta - \theta_0)^2 \end{aligned}$$

where $q_i$ is the ionic charge, $B_{ij}$ is the short-range Lenard-Jones repulsion parameter, $S_i$ and $D_i$ are the bond-valence and the bond-valence vector penalty energy scaling factors, respectively, and $k_i$ is the angle potential force constant. $V_i$ can be interpreted as the total number of bonds for each ion, and $\widehat{W}_i$ can be interpreted as the weighted vector sum of these bonds. More specifically, $V_i$ is defined as the sum of valence for all neighboring bonds:

$$V_i = \sum_j V_{ij}$$

where the valence of a bond is written as (*32-39*):

$$V_{ij} = (\frac{r_{ij}^0}{r_{ij}})^{C_{ij}}$$

where $i$ and $j$ denote different ions, $r_{ij}^0$ stands for the desired bond length between two ions, and $C_{ij}$ is Brown's empirical parameter. $\widehat{W}_i$ is defined as the sum of all bond-valence vectors (*32, 33, 38*):

$$\begin{aligned} \widehat{W}_i &= \sum_j \widehat{W}_{ij} \\ &= \sum_j V_{ij} \widehat{R}_{ij} \end{aligned}$$

where $\widehat{R}_{ij}$ is the unit vector pointing from ion $i$ to ion $j$. All parameters in the model are optimized by the simulated annealing method to fit energies and forces in DFT database. Detailed information about the bond-valence model can be found in reference (*32, 33*) and (*36*). In this study, Ti and O are represented by potential parameters from BaTiO$_3$ in a previous study assuming transferability (*40*), and parameters for Sr are directly trained from the Ba/SrTiO$_3$ alloy DFT database. A full list of parameters for SrTiO$_3$ is shown in Table S1.

MD simulation
Our simulations are performed in LAMMPS (*41*) for a 20×20×20 supercell in an isothermal-isobaric ensemble (*NPT*), where the temperature is controlled at 5 K by Nosé–Hoover thermostat, and the pressure is controlled at 1 atm by Parrinello-Rahman barostat (*42*). The thermal inertia parameter $M_s$ in Nosé–Hoover thermostat is chosen to be 1.0 amu. The profile of the applied electric field pulse is shown in Fig. S7. Forces on ions under the electric field are calculated from Born effective charge tensors ($Z^*_{Sr}$=2.56, $Z^*_{Ti}$=7.40, $Z^*_{O\perp}$=-2.08, and $Z^*_{O\parallel}$=-5.80) (*43*).

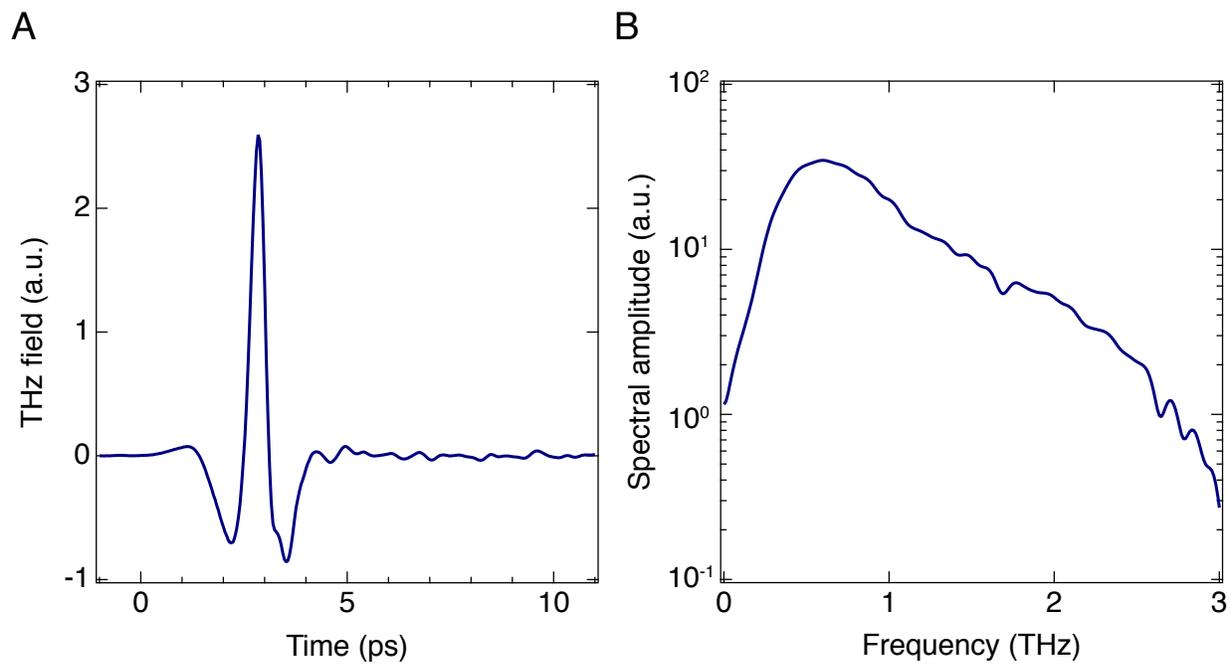

**Fig. S1.**
Measured THz field profile used in the experiments. **(A)** THz pump field time-domain signal measured by electro-optic sampling. **(B)** THz pump field spectrum on a logarithmic scale.

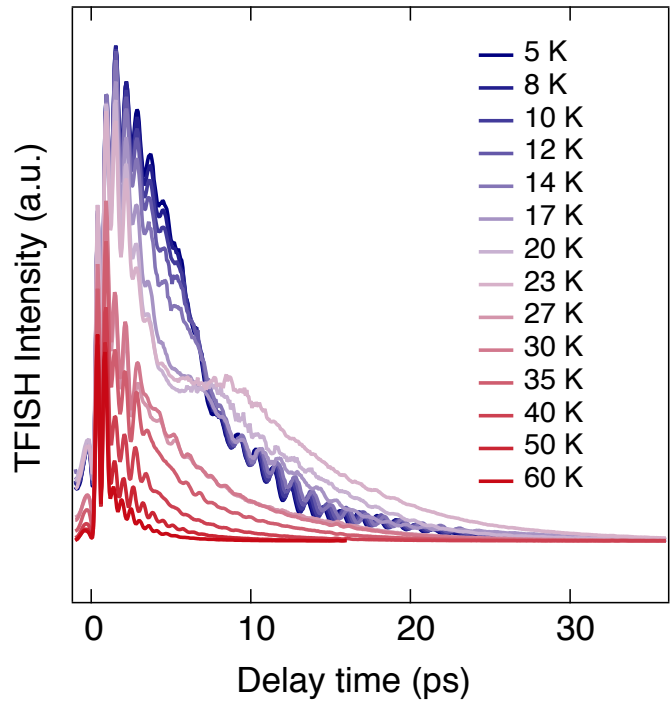 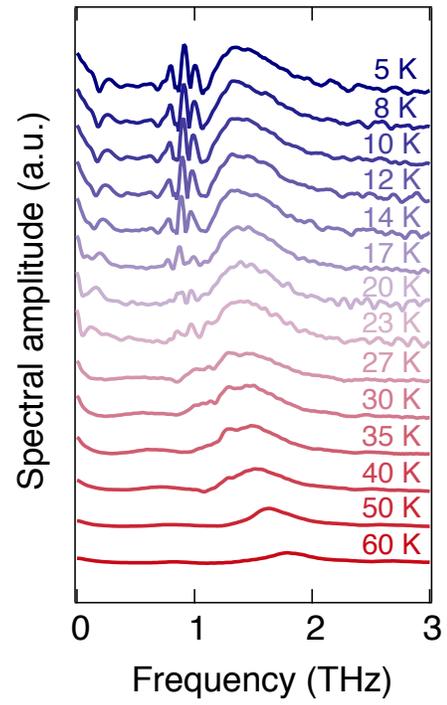

**Fig. S2.**
Complete TFISH temperature-dependent data. **(A)** TFISH time-domain temperature-dependent data with an incident THz field level of 550 kV/cm. **(B)** Fourier transforms of the TFISH data after exclusion of the electronic responses near $t = 0$ and numerical first differentiation.

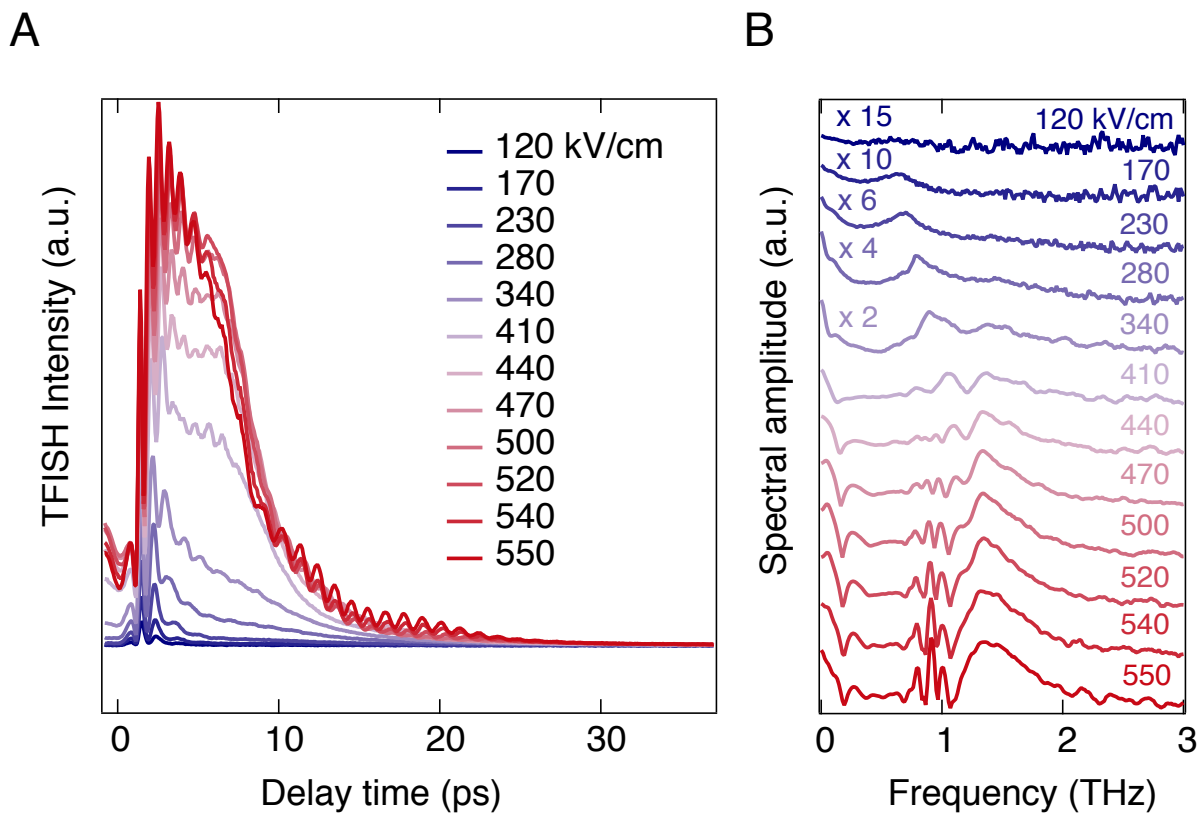

**Fig. S3.**
Complete TFISH THz field-strength-dependent data measured at 5 K. **(A)** TFISH time-domain field-dependent data. **(B)** Fourier transforms of the time-domain field-dependent data after exclusion of the electronic responses near $t = 0$ and numerical first differentiation.

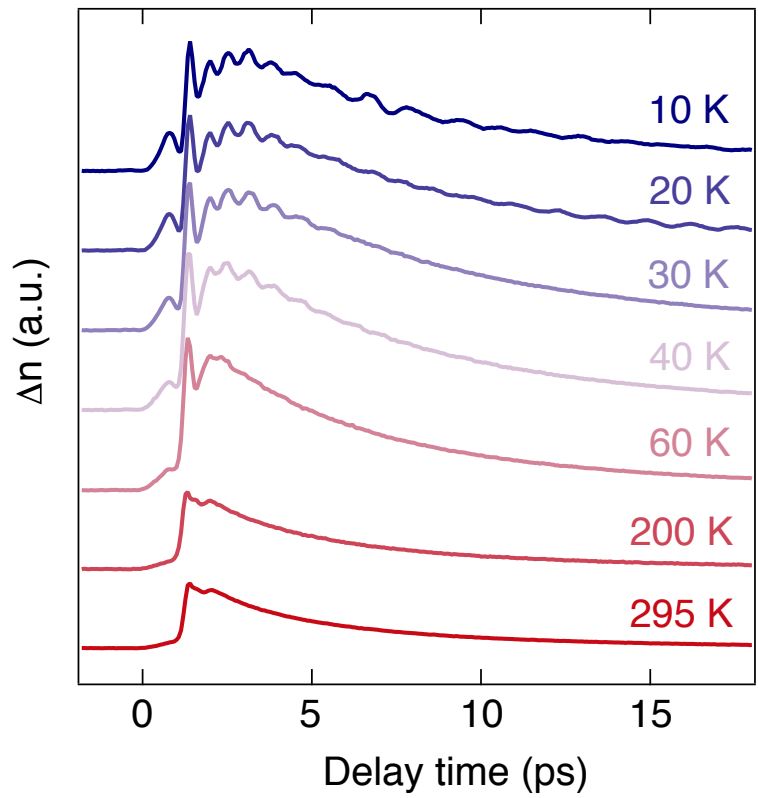

**Fig. S4.**
Complete TKE time-domain temperature-dependent data with an incident THz field level of 630 kV/cm.

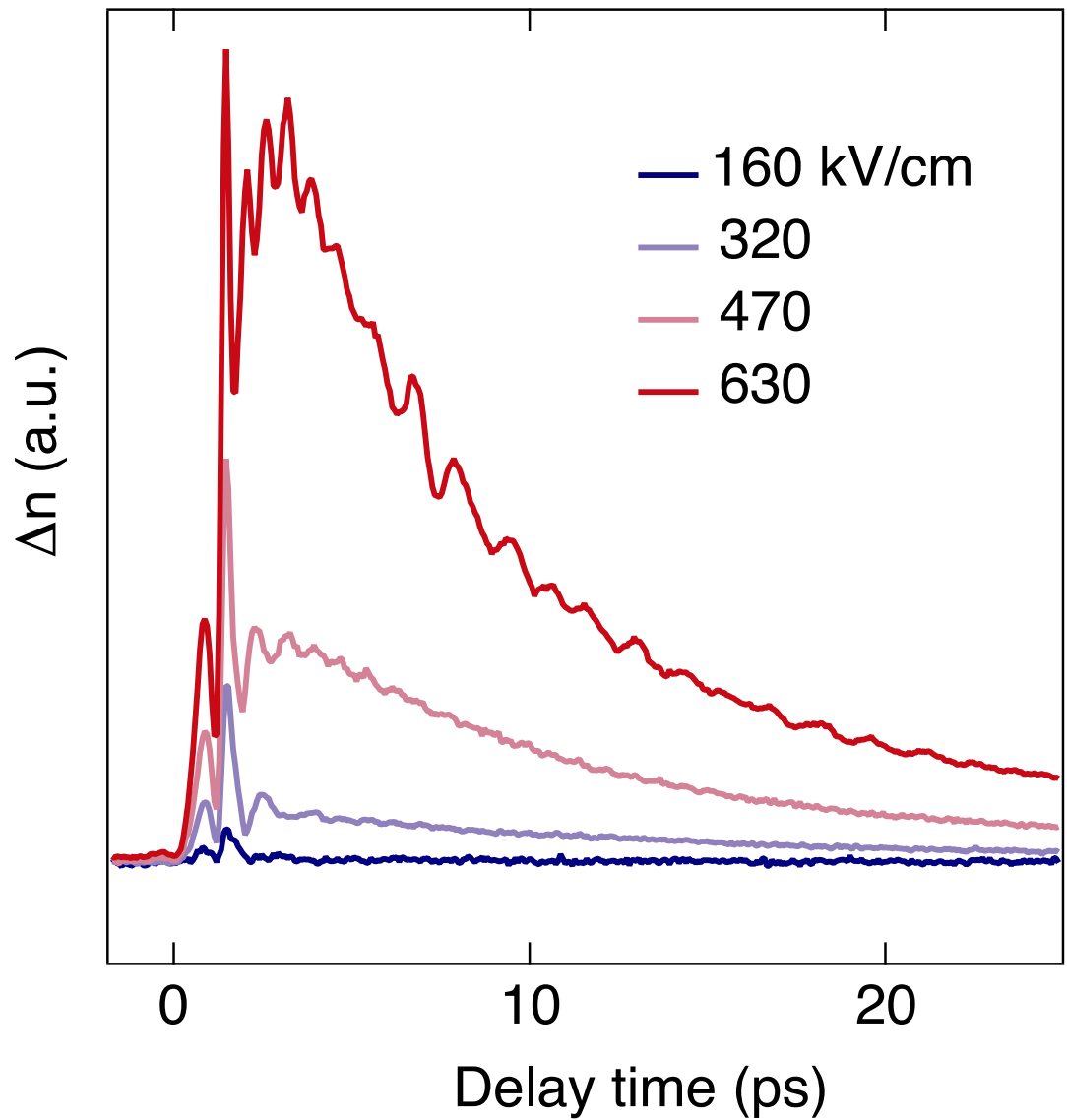

**Fig. S5.**
TKE time-domain field-dependence data measured at 10 K with a maximum incident THz field level of 630 kV/cm.

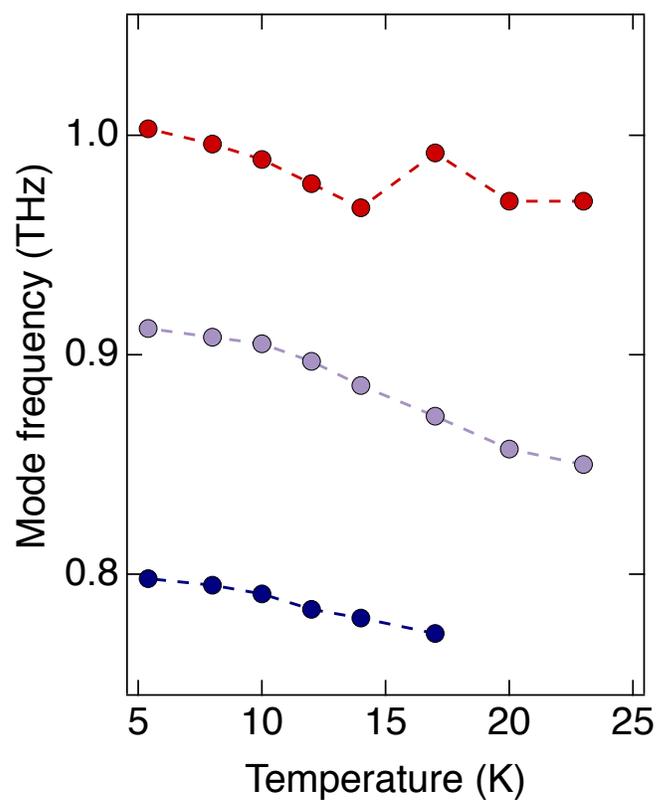

**Fig. S6.**
Frequency evolution of the three low-frequency modes that appear in the ferroelectric phase as a function of temperature. All three modes harden with decreasing temperature.

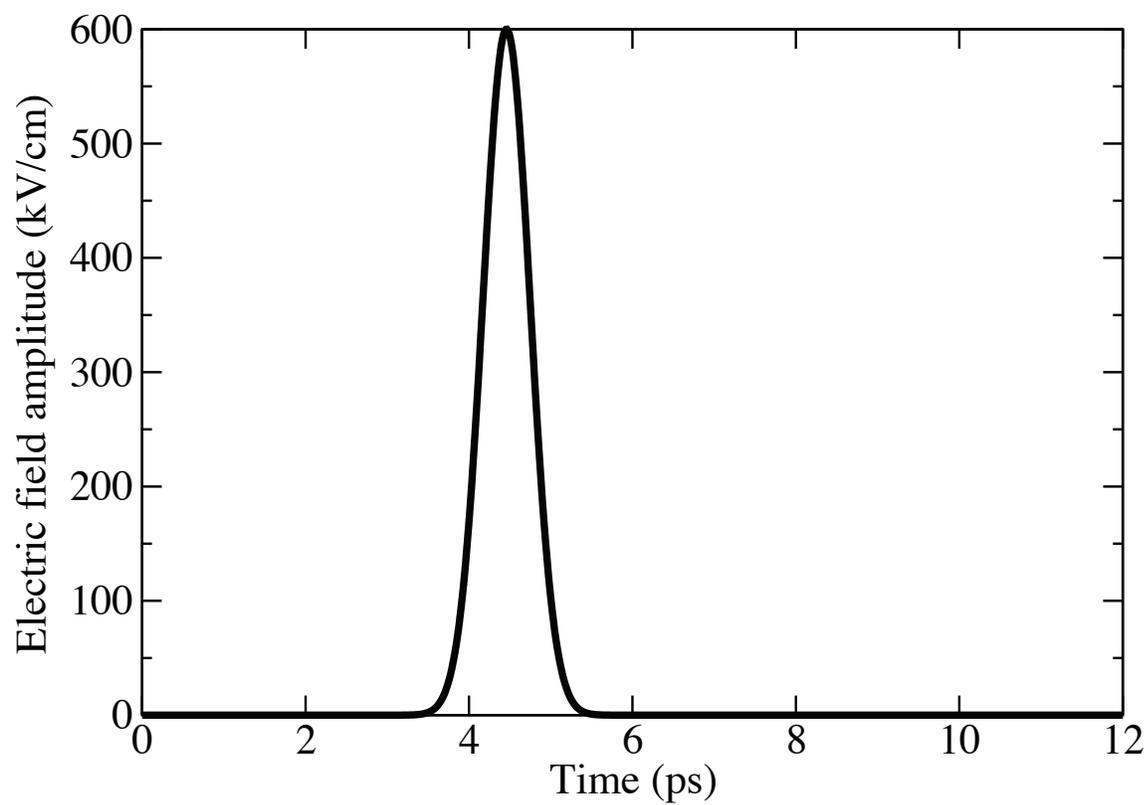

**Fig. S7.**
Profile of the applied electric field pulse used in MD simulations, for a field amplitude of 600 kV/cm.

|    | $r_{0,\beta O}$ | $C_{0,\beta O}$ | $q_\beta$ | $S_\beta$ (eV) | $D_\beta$ | \multicolumn{3}{c}{$B_{\beta\beta'}$ (Å)} | $V_{0,\beta}$ | $\widehat{W}_{0,\beta}$ |
|----|-----------------|-----------------|-----------|----------------|-----------|---------|---------|---------|---------------|--------------------------|
|    |                 |                 |           |                |           | Sr      | Ti      | O       |               |                          |
| Sr | 2.143           | 8.94            | 1.34730   | 0.63624        | 9.99121   | 0.38947 | 1.68014 | 1.96311 | 2.0           | 0.00000                  |
| Ti | 1.798           | 5.20            | 1.28905   | 0.16533        | 0.82484   |         | 2.73825 | 1.37741 | 4.0           | 0.39437                  |
| O  |                 |                 | -0.8787   | 0.93063        | 0.28006   |         |         | 1.99269 | 2.0           | 0.31651                  |

Table S1.
**Parameters for SrTiO$_3$ in bond-valence model**. The angle potential is set to zero.